\documentclass[letter,twocolumn]{jpsj2}
\usepackage{bm}
\usepackage{latexsym}
\usepackage{graphics}
\usepackage{graphicx}
\usepackage{amssymb}

\title{%
Detection of Neutron Scattering from Phase IV of Ce$_{0.7}$La$_{0.3}$B$_{6}$:\\
 A Confirmation of the Octupole Order
}

\author{%
Keitaro \textsc{Kuwahara}$^1$\thanks{E-mail address: kuwahara@phys.metro-u.ac.jp}, 
Kazuaki \textsc{Iwasa}$^2$,
Masahumi \textsc{Kohgi}$^1$, 
Naofumi \textsc{Aso}$^3$,\\
Masafumi \textsc{Sera}$^4$,
and Fumitoshi \textsc{Iga}$^4$
}

\inst{%
$^1$Department of Physics, Tokyo Metropolitan University, Tokyo 192-0397\\ 
$^2$Department of Physics, Tohoku University, Sendai 980-8578\\
$^3$Neutron Science Laboratory, Institute for Solid State Physics, University of Tokyo, Ibaraki 319-1106\\   
$^4$Department of Quantum Matter, ADSM, Hiroshima University,  Hiroshima 739-8530\\
}

\recdate{}

\abst{%
We have performed a single crystal neutron scattering experiment on Ce$_{0.7}$La$_{0.3}$B$_{6}$ to microscopically investigate the order parameter of phase~IV. 
Below the phase transition temperature 1.5~K of phase~IV, weak but distinct superlattice reflections at the scattering vector $\bm{\kappa}$ = ($\frac{h}{2}$,$\frac{h}{2}$,$\frac{l}{2}$)  ($h$, $l$ = odd number) have been observed for the first time by neutron scattering. The intensity of the superlattice reflections is stronger for high scattering vectors, which is quite different from the usual magnetic form factor of magnetic dipoles. This result directly evidences that the order parameter of phase~IV has a complex magnetization density, consistent with the recent experimental and theoretical prediction in which the order parameter is the magnetic octupoles $T^{\beta}$ with $\Gamma_5$ symmetry of the point group $O_h$. Neutron scattering experiments using short wavelength neutrons, as done in this study, could become a general method to study the high-rank multipoles in f electron systems.  
}

\kword{%
Ce$_{x}$La$_{1-x}$B$_{6}$, phase IV, neutron scattering, octupole order
}

\begin{document}
\sloppy
\maketitle

The importance of high-rank multipolar degrees of freedom of f electrons in strongly correlated electron systems has recently been widely recognized. A typical dense Kondo compound CeB$_6$ with a simple cubic crystal structure of space group $Pm{\bar 3}m$ is a well-known example where the importance was clarified experimentally and theoretically.
It shows the following two successive phase transitions: The first is from the paramagnetic phase (phase~I) to the antiferro-quadrupolar ordering phase characterized by the wave vector ${\mib k}_{\rm Q}$ = [$\frac{1}{2}$,$\frac{1}{2}$,$\frac{1}{2}$] at 3.3~K (phase~II), followed by the second transition to the antiferromagnetic ordering phase with a complex magnetic structure characterized by the four nonequivalent wave vectors at 2.3~K (phase~III),~\cite{effantin85} where the fifteen multipoles in the $\Gamma_8$ quartet crystal-field ground state of CeB$_6$ play an important role for these orderings.~\cite{sakai97,shiina97} By doping La into the Ce site in this system, a new phase called phase~IV appears below $T_{\rm IV}$ = 1.7~K and 1.5~K in Ce$_{x}$La$_{1-x}$B$_{6}$ for $x$ = 0.75 and 0.70, respectively.~\cite{hiroi97,tayama97,suzuki97,hiroi98,sakakibara02,morie04,akatsu03,suzuki05}  From the initial discovery of phase~IV,~\cite{hiroi97,tayama97,suzuki97}  its characteristic magnetic phase diagram and anomalous bulk properties, which show the isotropic cusp of  magnetization at $T_{\rm IV}$ and the strong elastic softening of $c_{44}$ $within$ phase~IV, suggested that the ordering of phase~IV must be different from any quadrupolar ordering. Since the $\Gamma_8$ quartet has three types of octupoles -- $T_{\rm xyz}$, $T^{\alpha}$, and $T^{\beta}$ -- in addition to dipoles and quadrupoles, it was argued that the magnetic octupoles are a possible candidate for the order parameter. Following these studies, the magnetization under uniaxial pressure, thermal expansion, and elastic constant measurements supported that the order parameter of phase~IV is $T^{\beta}$ octupoles with $\Gamma_5$ symmetry of the point group $O_h$;~\cite{sakakibara02,morie04,akatsu03,suzuki05} it is consistent with the existence of internal magnetic fields detected by NMR~\cite{magishi02} and $\mu$SR~\cite{takigawa02,schenck07} as well as a theoretical model~\cite{kubo04}.  Furthermore, evidence of the antiferro-octupolar ordering of $T^{\beta}$ with the same wave vector as ${\mib k}_{\rm Q}$  has recently been reported by the resonant X-ray scattering experiment and its detailed analysis,~\cite{mannix05,kusunose05}
although further studies are necessary for understanding the overall nature of phase~IV in Ce$_{x}$La$_{1-x}$B$_{6}$.~\cite{kondo07}

In principle, such time-reversal-symmetry-breaking high-rank multipoles can be detected by neutron scattering, because a neutron interacts with electrons through magnetic interactions.~\cite{squires78,lovesey84} Moreover, this probe has the advantage of being able to get  direct information about  the magnetization density of high-rank multipoles from the magnetic form factor. Notwithstanding this expectation, no significant magnetic Bragg peaks were observed for phase~IV within experimental accuracy in our previous neutron scattering experiments below $\frac{|\kappa|}{4 \pi}$ = $\frac{\sin \theta}{\lambda}$ = 0.4\AA$^{-1}$, where ${\mib \kappa}$, $\theta$, and $\lambda$ are the scattering vector, Bragg angle, and neutron wavelength, respectively.~\cite{kohgi00,iwasa03,fischer05} Although the contribution from magnetic octupoles to the neutron scattering cross section  may be small, it must be finite and the scattering intensity in the high ${\mib \kappa}$ vector region is expected to be stronger than that in the low ${\mib \kappa}$ vector,~\cite{lovesey84,shiina07} because magnetic octupoles must have a complex magnetization density with no spatially uniform magnetization. Furthermore, a recent theoretical calculation of the magnetic form factor of octupoles predicts a detectable scattering intensity magnitude.~\cite{shiina07}

In this letter, we report the recent results of the neutron scattering experiment on a Ce$_{0.7}$La$_{0.3}$B$_{6}$ single crystal, focusing attention on superlattice reflections in the high scattering vector region. We have succeeded in detecting weak but distinct superlattice reflections from phase~IV by neutron scattering for the first time. The ${\mib \kappa}$ dependence of the magnetic form factor in the superlattice spots directly evidences that  the order parameter of phase~IV  has a complex structure of magnetization density, consistent with the theoretical and experimental prediction that the order parameter  is the magnetic octupoles $T^{\beta}$ with $\Gamma_5$ symmetry of $O_h$.

A large single crystal of Ce$_{0.7}$La$_{0.3}$B$_{6}$ was grown by the floating zone method, using 99.52\% enriched $^{11}$B to avoid the large neutron absorption due to $^{10}$B. The bulk properties of the single crystal were checked  by electrical resistivity and magnetization measurements. The sample is cylindrical in shape with 
a diameter and length of 4.4~mm and 14~mm, respectively. 
The cylinder axis is nearly parallel to the [010] direction. The neutron scattering experiment was performed on the thermal neutron triple-axis spectrometer TOPAN (6G) at the JRR-3M reactor in the Japan Atomic Energy Agency. The sample was mounted in the mixing chamber of a $^3$He-$^4$He dilution refrigerator with a superconducting magnet. Magnetic fields were applied along  the [${\bar 1}$,1,0] direction, normal to the ($h$,$h$,$l$) scattering plane. Incident neutrons with the short wavelength $\lambda$ = 1.4133\AA\   were selected by a pyrolytic graphite (PG) monochromator in order to search for superlattice reflections in the high scattering vector region. The triple-axis mode was used with the collimation open-60$'$-60$'$-60$'$ and double PG filters to get a better signal-to-noise ratio. In this experimental setup, the mosaicity of the sample is 0.36$^\circ$ full width at half maximum (FWHM), reflecting the good quality of the single crystal.

For determining of the magnetic form factor, some corrections are needed. The Lorentz factor and absorption factor corrections were made for the observed nuclear and superlattice reflections. The former was represented by  $1/\sin 2\theta$ in the present geometry of scans. The latter was numerically calculated by the Fortran program by taking the approximate shape of the sample into account. The change in the intensity due to the absorption correction is less than 16\%. To obtain the absolute value of the magnetic form factor, information about the normalization factor  between the integrated intensity of fundamental nuclear Bragg reflections and the intensity calculated from the nuclear structure factors is also needed. For calculating the nuclear structure factors, the most reliable site parameter of B determined by the previous powder neutron experiment on Ce$_{0.75}$La$_{0.25}$B$_{6}$~\cite{fischer05} was used. However, the integrated intensity of several nuclear Bragg reflections is not proportional to the calculated value,  especially for strong Bragg reflections. This deviation may be caused by the unavoidable extinction effects~\cite{givord03}, which strongly influence the Bragg intensity  in experiments using a large single crystal with a small mosaicity as used in this study. It is difficult to correct for the influence of the extinction effects on the nuclear Bragg intensity in this case. Thus, the normalization factor was estimated by using the intensity of three relatively weak Bragg reflections (2,2,0), (1,1,0), and (0,0,1).

\begin{figure}[rt]

\begin{center}

\includegraphics[width=6.5cm]{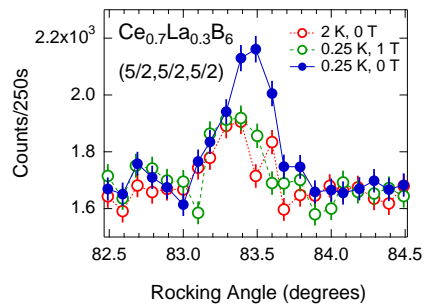}

\end{center}

\caption{(Color online) Scattering patterns of rocking curves  at the scattering vector ${\mib \kappa}$ = ($\frac{5}{2}$,$\frac{5}{2}$,$\frac{5}{2}$)  in  phase~I (at temperature $T$ = 2~K and magnetic field $B$ = 0~T), phase~III ($T$ = 0.25~K, $B$ = 1~T), and phase~IV ($T$ = 0.25~K, $B$ = 0~T) in Ce$_{0.7}$La$_{0.3}$B$_{6}$. The magnetic field is applied along the [${\bar 1}$,1,0] direction.
}

\label{f1}

\end{figure}

\begin{figure}[rt]

\begin{center}

\includegraphics[width=5.5cm]{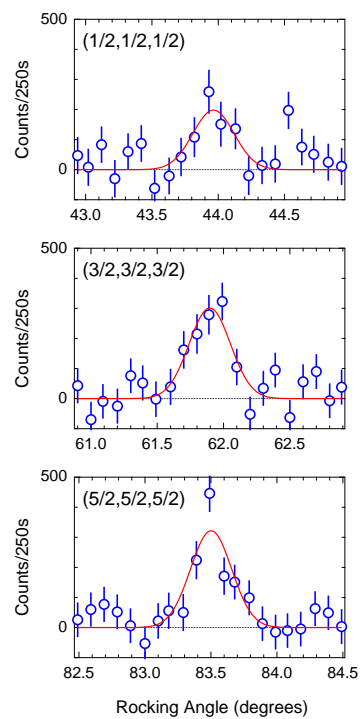}

\end{center}

\caption{(Color online) Difference diffraction patterns between  0.25~K and  2~K under a zero magnetic field at ${\mib \kappa}$ = ($\frac{h}{2}$,$\frac{h}{2}$,$\frac{l}{2}$) along the [1,1,1]  direction in Ce$_{0.7}$La$_{0.3}$B$_{6}$. The lines are  Gaussian fits.}

\label{f2}

\end{figure}

\begin{figure}[]

\begin{center}

\includegraphics[width=6.5cm]{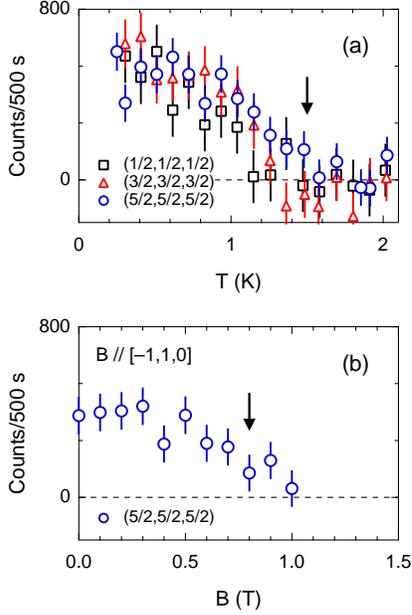}

\end{center}

\caption{(Color online) (a) Temperature  dependence of the intensity at the peak positions of ${\mib \kappa}$ = ($\frac{1}{2}$,$\frac{1}{2}$,$\frac{1}{2}$), ($\frac{3}{2}$,$\frac{3}{2}$,$\frac{3}{2}$), and ($\frac{5}{2}$,$\frac{5}{2}$,$\frac{5}{2}$) in a zero magnetic field, (b) magnetic field dependence along the [${\bar 1}$,1,0] direction of the intensity at the peak position of ${\mib \kappa}$ = ($\frac{5}{2}$,$\frac{5}{2}$,$\frac{5}{2}$) at 0.25~K, where the background intensity is subtracted.
The arrows indicate the phase boundary of phase~IV reported by the bulk measurements.}

\label{f3}

\end{figure}

At the lowest temperature 0.25 K under a zero magnetic field in phase~IV, we have observed weak superlattice reflections characterized by the wave vector [$\frac{1}{2}$,$\frac{1}{2}$,$\frac{1}{2}$] for the first time. This wave vector is the same as determined by the recent resonant X-ray scattering experiment.~\cite{mannix05} Figure~1 shows an example of the scattering patterns of rocking curves in phases I, III, and IV. The enhancement of the scattering intensity in phase~IV has been clearly observed. The peaks seen in phases I and III are the extrinsic higher-order contamination from the strong (5,5,5) nuclear Bragg reflection. It should be noted that the observed distinct enhancement of intensity  does not come from the extinction effects, which produce an increase of about 1\% in the nuclear Bragg reflections in phase~IV~\cite{kohgi00}, because the extinction effects do not affect such a weak signal and the enhancement of the intensity of the superlattice reflections is at least an order of magnitude larger than that of the nuclear Bragg reflections. Therefore, we believe that the observed peaks are an intrinsic signal from the order parameter of phase~IV. Figure~2 shows the difference diffraction pattern between 0.25~K and 2~K under a zero magnetic field at ${\mib \kappa}$ = ($\frac{h}{2}$,$\frac{h}{2}$,$\frac{l}{2}$) along the [1,1,1] direction with threefold symmetry. The lines are Gaussian fits.  In the fitting procedure, the width of the Gaussian profile function is fixed to that of the fundamental Bragg reflections 0.36$^\circ$ (FWHM). We also observed several superlattice reflections along other directions of ${\mib \kappa}$, such as ($\frac{5}{2}$,$\frac{5}{2}$,$\frac{1}{2}$), but reliable data with sufficient statistical accuracy have not been obtained because of the very weak signal.  
Taking the Lorentz factor correction into account, we note that the integrated intensity at ${\mib \kappa}$ = ($\frac{3}{2}$,$\frac{3}{2}$,$\frac{3}{2}$) and ($\frac{5}{2}$,$\frac{5}{2}$,$\frac{5}{2}$) is stronger than that at ${\mib \kappa}$ = ($\frac{1}{2}$,$\frac{1}{2}$,$\frac{1}{2}$); this is different from the usual magnetic form factor, as discussed later.

To confirm that the observed signal comes entirely from the order parameter of phase~IV, we measured the temperature and magnetic field dependences of the peak intensity of the superlattice reflections, as shown in Figs.~3(a) and 3(b). In the temperature dependence under a zero magnetic field, the peaks  at ${\mib \kappa}$ = ($\frac{1}{2}$,$\frac{1}{2}$,$\frac{1}{2}$), ($\frac{3}{2}$,$\frac{3}{2}$,$\frac{3}{2}$), and ($\frac{5}{2}$,$\frac{5}{2}$,$\frac{5}{2}$) develop below $T_{\rm IV}$ = 1.4~K with decreasing temperatures. In the magnetic field dependence along the [${\bar 1}$,1,0]  direction at  0.25~K, the  peak intensity  at ${\mib \kappa}$ =  ($\frac{5}{2}$,$\frac{5}{2}$,$\frac{5}{2}$) seems to disappear above about 1~T, which agrees with the phase boundary of phase~IV under magnetic fields along the [${\bar 1}$,1,0] direction~\cite{hiroi98}.
These results clearly show that the observed weak superlattice peaks come from the order parameter of phase~IV.

In general, the neutron can be scattered by time-reversal-symmetry-breaking  multipoles due to an electron with spin $\bm{s}$ and momentum $\bm{p}$ through the magnetic interaction, because the neutron has the magnetic dipole moment. The cross section of the single electron is proportional to $|\langle\lambda|\bm{Q}_{\perp}|\lambda\rangle|^2$, where $|\lambda\rangle$ is a state of the electron. The scattering operator $\bm{Q}_{\perp}$ is defined by
\begin{equation}
\bm{Q}_{\perp} =  \exp(i\bm{\kappa}\cdot\bm{r}) \left\{ \tilde{\bm{\kappa}} \times (\bm{s} \times \tilde{\bm{\kappa}}) - \frac{i}{\hbar |\bm{\kappa}|} \tilde{\bm{\kappa}} \times \bm{p} \right\},
\end{equation}
where $\bm{r}$ denotes the position of the electron and $\tilde{\bm{\kappa}} = \bm{\kappa}/|\bm{\kappa}|$.~\cite{lovesey84} $\bm{Q}_{\perp}$ is related to  the Fourier transform of the magnetization density $\bm{M}(\bm{\kappa})$ of the electron as follows:
\begin{equation}
2  \bm{Q}_{\perp} \mu_{\rm B} = - \tilde{\bm{\kappa}} \times (\bm{M}(\bm{\kappa}) \times \tilde{\bm{\kappa}}).
\end{equation}
The expectation value of $2 \bm{Q}_{\perp}$ is the magnetic form factor. Since the magnetization density of octupoles is completely different from that of dipoles, as seen in eq.~(2), 
their magnetic form factors show a qualitatively different behavior in the $\bm{\kappa}$ dependence. In the dipolar case, the form factor decreases with increasing $\bm{\kappa}$, such as that reported in the pure system CeB$_6$~\cite{givord03}. On the other hand, in the octupolar case,  the form factor is zero at $\bm{\kappa}$ = 0 and has a maximum at a finite $\bm{\kappa}$, reflecting its complex magnetization density with no spatially uniform magnetization. The characteristic behavior of the form factor of octupoles is given by the detailed theoretical calculation of $2 |\langle\lambda|\bm{Q}_{\perp}|\lambda\rangle|$, taking the orbital contribution of the second term of eq.~(1) correctly, in which the atomic wave function diagonal for the magnetic octuples $T^{\beta}$ in the $\Gamma_8$ quartet is assumed.~\cite{shiina07} Therefore, we can distinguish whether the observed superlattice reflections  originate  from dipoles or octupoles by the $\bm{\kappa}$ dependence of the magnetic form factor.

\begin{figure}[]

\begin{center}

\includegraphics[width=6cm]{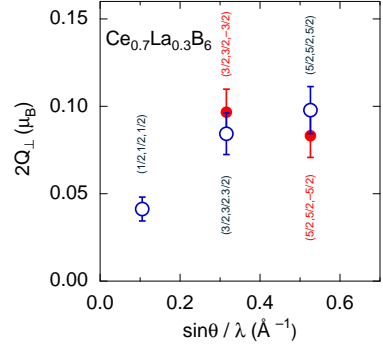}

\end{center}

\caption{(Color online) Magnetic form factor at the superlattice spots along the [1,1,1] (open circles) and [1,1,${\bar 1}$]  (filled circles) directions in phase~IV of Ce$_{0.7}$La$_{0.3}$B$_{6}$.}

\label{f4}

\end{figure}

Figure~4 shows the magnetic form factor at the superlattice spots along the [1,1,1] and [1,1,${\bar 1}$] directions obtained by using the integrated intensity of the observed superlattice reflections and by making the corrections. The absolute value of the form factor is slightly lesser than 0.1$\mu_{\rm B}$.  This magnitude especially at the smallest scattering vector ${\mib \kappa}$ = ($\frac{1}{2}$,$\frac{1}{2}$,$\frac{1}{2}$) with $\frac{\sin \theta}{\lambda}$ = 0.1\AA$^{-1}$, which is expected to be the largest scattering intensity in the usual magnetic neutron scattering, is quite small; it is comparable to the magnitude of tiny magnetic dipole moments reported in some heavy electron systems.
This smallness must be the reason why the magnetic reflections were not previously detected even by the highest intensity powder neutron diffraction.~\cite{fischer05} The form factor in Fig.~4 is clearly strong for high scattering vectors. This cannot be explained by the usual antiferromagnetic ordering even by considering any conceivable magnetic structure as well as any domain distribution because the data in Fig.~4 are the form factors at the superlattice spots along the same direction. Therefore, this unusual ${\mib \kappa}$ dependence of the form factor directly evidences that the order parameter has a magnetization density different from ordinary dipole orderings. This result  qualitatively agrees with the theoretical calculation considering an average of four domains of the order parameter $T^{\beta}$. Furthermore, from the selection rule of the cross section based on the symmetry classification of octupolar scattering for the three possible octupoles $T_{xyz}$,  $T^{\alpha}$, and  $T^{\beta}$,~\cite{shiina07}  the present result can rule out the possibility of $T_{xyz}$ with $\Gamma_2$ symmetry because the superlattice reflections along the [1,1,1] direction with threefold symmetry have been observed. $T^{\alpha}$ with $\Gamma_4$ symmetry is also unlikely to explain the observed ${\mib \kappa}$ dependence because magnetic dipoles with the same $\Gamma_4$ symmetry as $T^{\alpha}$ are expected to be mixed. Therefore, the present result strongly indicates that the order parameter of phase~IV is the magnetic octuples $T^{\beta}$ with $\Gamma_5$ symmetry. For a quantitative comparison between the experimental form factor and the theoretical calculation including the anisotropy of the form factor, detailed corrections of the extinction effects and data at more superlattice reflection points as well as more statistical accuracy are necessary. 

It still remains possible that such a ${\mib \kappa}$ dependence of intensity might arise from lattice distortions, since the scattering intensity due to lattice distortions also increases with increasing ${\mib \kappa}$ in proportion to the square of ${\mib \kappa}$. This possibility cannot be completely ruled out only on the basis of the present neutron scattering experiment, but it should be noted that no superlattice reflection due to lattice distortions has been observed by X-ray scattering, which is the most powerful probe for detecting lattice distortions. This fact as well as the existence of the internal fields detected by NMR~\cite{magishi02} and $\mu$SR~\cite{takigawa02,schenck07} support that the observed superlattice reflections are magnetic. To definitively confirm this, we will perform a polarized neutron scattering experiment soon. 

In summary, weak but distinct superlattice reflections at $\bm{\kappa}$ = ($\frac{h}{2}$,$\frac{h}{2}$,$\frac{l}{2}$)  ($h$, $l$ = odd number) from  phase~IV of Ce$_{0.7}$La$_{0.3}$B$_{6}$ have been observed for the first time by elastic neutron scattering using a large single crystal. The intensity of the superlattice reflections is stronger for high scattering vectors. This unusual $\bm{\kappa}$ dependence of the intensity evidences that the order parameter of phase~IV has a complex magnetization density, consistent with the recent experimental and theoretical prediction in which the order parameter is the magnetic octupoles $T^{\beta}$ with $\Gamma_5$ symmetry of $O_h$. Neutron scattering experiments using short wavelength neutrons, as done in this study, could become a general method to study the high-rank multipoles in f electron systems. Further neutron scattering measurements, including experiments under uniaxial stress, are planned to clarify the detailed nature of the order parameter of phase~IV.

We wish to thank R.\ Shiina, O.\ Sakai, H.\ Shiba, H.\ Kusunose, Y.\ Kuramoto, J.-M.\ Mignot, H.\ Kadowaki, Y.\ Tanaka, and K.\ Katsumata for discussions and comments. 
This work was supported by Grants-in-Aid for Young Scientists (B) (No.\ 17740236) and Scientific Research Priority Area {\lq}{\lq}Skutterudites{\rq}{\rq} (No.\ 15072206) from the Ministry of Education, Culture, Sports, Science and Technology of Japan.

\end{document}